\documentstyle[prl,aps,epsf,floats]{revtex}

\begin{document}
\draft

\twocolumn[\hsize\textwidth\columnwidth\hsize\csname @twocolumnfalse\endcsname

\title{Highly polarized injection luminescence in
forward-biased ferromagnetic-semiconductor junctions at low spin polarization
 of current}
\author{A.M. Bratkovsky and V.V. Osipov}
\address{
Hewlett-Packard Laboratories, 1501 Page Mill Road, 1L, Palo Alto, CA 94304
}
\date{July 12, 2004}
\maketitle

\begin{abstract}
We consider electron tunneling  from a nonmagnetic $n$-type semiconductor ($n$-S) 
into a ferromagnet (FM) through a very thin forward-biased Schottky barrier resulting
in efficient extraction of electron spin from a thin $n$-S layer near FM-S interface 
at low spin polarization of the current.  We show that this effect can be used for 
an efficient polarization radiation source in a heterostructure where
the accumulated spin polarized 
electrons are injected from $n$-S and recombine with holes in a quantum well.
The radiation polarization depends on a bias voltage applied to the FM-S junction. 	
\end{abstract}

\vskip2pc]

\narrowtext

The desire of building scaleable devices using an electron spin created a
new field of spintronics \cite{Wolf}. Some spintronics effects, like a giant
magnetoresistance of multilayers and tunnel structures \cite{GMR,Slon,Brat},
already found many applications. Of particular interest is an injection of
spin-polarized electrons into nonmagnetic semiconductors (NS) because of a
long spin-relaxation time \cite{Wolf,Aw} and a prospect of using this
phenomenon for ultrafast spintronic devices and possibly quantum computers 
\cite{Wolf}. Different ferromagnet-semiconductor-ferromagnet (FM-S-FM)
structures have been suggested including those where electron spin is
affected by an applied electric field \cite{Datta}, an external magnetic
field \cite{BO}, or a magnetic field produced by a nanowire current \cite{OB}%
. All these devices are controllable spin valves where one of FM-S junctions
acts as a spin source and another one as a spin filter. The latter
efficiently transmits electrons with a certain spin projection and reflects
electrons with the opposite spin. The spin polarization of photoexcited
electrons in NS, some effects and devices due to spin reflection off a
ferromagnet have been studied in Refs. \cite{Ciuti}. Spin injection from
magnetic semiconductor in NS has been reported in Refs. \cite{MSemi}. Spin
accumulation and extraction from NS was predicted in Ref. \cite{Zut} in
forward-biased p-n junctions containing magnetic semiconductors and observed
recently in MnAs/GaAs junction \cite{Step}. Spin injection from FM into NS
was demonstrated in Refs. \cite{Ferro,Jonk}. Optimal conditions of the spin
injection\ from FM into NS have been discussed in Refs. \cite
{BO,OB,Aron,Son,Flat,Alb,OBO,BOB}.

The injection of spin polarized electrons from FM into NS is determined by a
current in a reverse-biased FM-S Schottky junction. Such a current is
negligible due to a high Schottky barrier usually forming at FM-S interface 
\cite{sze}. Therefore, an ultrathin heavily doped interfacial semiconductor
layer ($\delta -$doped layer) is used to thin down the barrier considerably
and increase the current \cite{Jonk,BO,OB,Alb,OBO}. The conditions for the
most efficient spin injection in a reverse-biased FM-S junction have been
studied in Refs. \cite{Alb,BO,OB,OBO}. In forward-biased FM-S Schottky
junctions, thermionic emission current reaches large values at a bias
voltage $V$ close to $\Delta /q$, where $\Delta $ is the barrier height and $%
q$ the elementary charge. Realization of the spin injection due to such a
current is problematic since electrons in FM with an energy $E=F+\Delta $
are weakly spin polarized ($F$ is the Fermi level in FM). Recently we
considered a forward-biased FM-S junction with a $\delta -$doped layer and
showed \cite{BOB} that tunnelling of electrons from a NS into a FM through
the $\delta -$doped layer can result in efficient extraction of electrons
with a certain spin and accumulation of electrons with the opposite spin in
NS near the interface.

In this paper we consider a structure containing a FM-S junction with the $%
\delta -$doped layer and a double $p-n^{\prime }-n$ heterostructure where $%
n^{\prime }$-region made from narrower gap semiconductor, Fig. 1. We show
that the following effects can be realized in the structure when both FM-S
junction and the heterostructure are biased in forward direction and
electrons are injected from $n$-semiconductor region into FM and $p$-region.
Due to a spin selection property of FM-S junction \cite{BOB}, spin polarized
electrons appear in $n$-region with a spatial extent $L\lesssim L_{s}$ near
the FM-S interface, where $L_{s}$\ is the spin diffusion length in NS. When
thickness of\ the $n$-region $w$ is smaller than $L,$ the spin polarized
electron from the $n$-region and holes\ from $p$-region are injected and
accumulated in a thin narrow-gap $n^{\prime }$-region (quantum well) where
they recombine\ and emit polarized photons. The degree of the polarization $%
P $ depends on the bias voltage of FM-S junction, $V$, since different $V$
correspond to electron states in FM\ with different polarization [see curve
for $x<0$ in Fig. 1(c)]. Thus, the structure allows to study the dependence $%
P(V)$ and find the optimal bias $V$ such that the polarization of spin and
radiation is maximal.

\begin{figure}[t]
\epsfxsize=3.2in 
\epsffile{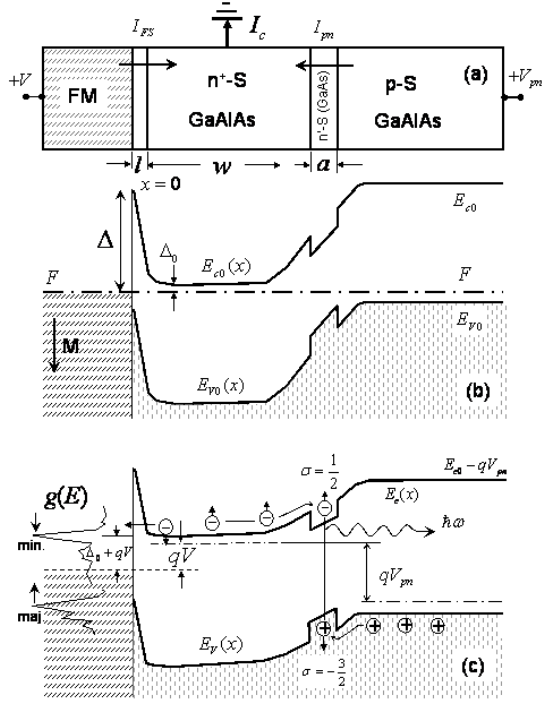}
\caption{Schematic (a) of structure and the band diagram of polarized photon
source containing FM-S junction with $\protect\delta -$doped layer and a
double $n^{+}-n^{\prime }-p$ heterostructure without (b) and under the bias
voltage $V$. Minority spin electrons are extracted from $n^{+}-S$
semiconductor layer and the remaining (majority)\ electrons are recombined
in $n^{\prime }-S$ quantum well. $F$ is the Fermi level, $\Delta $ the
height and $l$ the thickness of the $\protect\delta -$doped layer, $\Delta
_{0}$ the height of a barrier in the $n-$type semiconductor, $E_{c}(x)$ the
bottom of conduction band in the semiconductor. The spin density of states
is shown at $x<0,$ with a high peak in minority states at $E=F+\Delta _{0},$
typical of elemental Ni, as an example.}
\label{fig:fig1}
\end{figure}

Let us now consider these effects in greater detail. The spatial
distribution of a spin density in a semiconductor is determined by the
continuity equation \cite{Aron,Flat} 
\begin{equation}
\nabla J_{\sigma }=q\delta n_{\sigma }/\tau _{s},  \label{Kin}
\end{equation}
where $J_{\sigma }$ is the current of electrons with spin $\sigma =\uparrow
,\downarrow $, $\delta n_{\sigma }=n_{\sigma }-n/2$, $n$ and $\tau _{s}$ are
the density and the spin-relaxation time of electrons in the $n$%
-semiconductor region. In Ref. \cite{BOB} we showed that a tunneling current
in a forward-biased FM-S junction with the $\delta -$doped layer is

\begin{eqnarray}
J &=&J_{m}\gamma ,\text{ }\gamma =1-e^{-qV_{FS}/T},  \label{TC} \\
J_{m} &=&1.6(1-P_{F}^{2})(d_{\uparrow }+d_{\downarrow })nqv_{T}D_{t},
\label{Jm}
\end{eqnarray}
the current of electrons with spin $\sigma =\uparrow $ near the interface,
and $J_{\uparrow 0}\equiv J_{\uparrow }(0)$, given by 
\begin{equation}
J_{\uparrow 0}=\frac{J}{2}\frac{\left( 1+P_{F}\right) \left( \gamma
+P_{n0}\right) }{\gamma +P_{F}P_{n0}},  \label{J0}
\end{equation}
and the distribution of $\delta n_{\uparrow }(x)=-\delta n_{\downarrow }(x)$
is characterized by a length $L=(1/2)(\sqrt{L_{E}^{2}+4L_{S}^{2}}-L_{E})$,
where $L_{s}=\sqrt{D\tau _{s}}$ and $L_{E}=\mu \tau _{s}E=$ $L_{s}J/J_{s}$
are the diffusion and the drift lengths of electron spin in the NS,
respectively \cite{Aron,Flat}; $D$ and $\mu $ are the diffusion constant and
the mobility of electrons, and $E=J/q\mu n$ is the electric field in S. We
have introduced $J_{S}\equiv qDn/L_{s}$, the tunneling transparency of the
modified FM-S junction $D_{t}\simeq 1$ \cite{BO,OB,BOB}, the spin
polarization of nonequilibrium electrons near interface that defines the
degree of spin {\em extraction} from S, $P_{n0}=[n_{\uparrow
}(0)-n_{\downarrow }(0)]/n=2\delta n_{\uparrow 0}/n$, the spin factor $%
d_{\sigma }=v_{T}v_{\sigma 0}(v_{t0}^{2}+v_{\sigma 0}^{2})^{-1}$, and the
effective spin polarization characterizing spin selection of the modified
FM-S junction \cite{BO,OB,BOB}

\begin{equation}
P_{F}=\frac{d_{\uparrow }-d_{\downarrow }}{d_{\uparrow }+d_{\downarrow }}=%
\frac{(v_{\uparrow 0}-v_{\downarrow 0})(v_{t0}^{2}-v_{\uparrow
0}v_{\downarrow 0})}{(v_{\uparrow 0}+v_{\downarrow
0})(v_{t0}^{2}+v_{\uparrow 0}v_{\downarrow 0})}.  \label{PF}
\end{equation}
Here $v_{\sigma 0}=v_{\sigma }(F+\Delta _{0}+qV_{FS})$ is the velocity of
electron with spin $\sigma $ in the FM, $v_{t0}=\sqrt{2(\Delta -\Delta
_{0}-qV_{FS})/m_{\ast }}$ the tunneling ``velocity'' in the NS, $v_{T}\equiv 
\sqrt{3T/m_{\ast }}$ and $m_{\ast }$ are the thermal velocity and the
effective mass of electrons\ in\ NS, $V_{FS}$ the voltage drop across the
FM-S interface, $\Delta $ is height of potential barrier at FM-S interface, $%
\Delta _{0}=E_{c0}-F$, $E_{c0}$ the bottom of conduction band in $n$-region
in equilibrium, Fig. 1(b).

The conditions for maximal polarization are obtained as follows. When the
thickness of $n$-region is $w<L,$ we can assume that $\delta n_{\uparrow
}(x)\simeq \delta n_{\uparrow 0}$ and $P_{n}\simeq P_{n0}$. In this case,
integrating Eq. (\ref{Kin}) over the volume of the $n$-semiconductor region
(with area $S$ and thickness $w$), we obtain 
\begin{equation}
I_{\uparrow FS}+I_{\uparrow pn}-I_{c\uparrow }=q\delta n_{\uparrow 0}wS/\tau
_{s}=P_{n0}I_{S}w/2L_{s}.  \label{I-I}
\end{equation}
Here $I_{S}=J_{S}S$; $I_{\uparrow FS}=J_{\uparrow 0}S$ and $I_{\uparrow pn}$
are the electron currents with spin $\sigma =\uparrow $ flowing into the $n$%
-region from FM and the $p$-region, respectively; $I_{\uparrow c}$ is the
spin current out of the $n$-region in a contact, Fig. 1(a). The current $%
I_{\uparrow pn}$ is determined by injection of electrons with $\sigma
=\uparrow $ from the $n$-region into the p-region, equal to $I_{\uparrow
pn}=I_{pn}n_{\uparrow 0}/n=I_{pn}(1+P_{n0})/2$, where $I_{pn}$ is the total
current in the $p-n$ junction. The current of metal contact $I_{c}$ is not
spin polarized, hence $I_{c\uparrow }=(I_{pn}+I_{FS})/2$, where $I_{FS}$ is
the total current in FM-S junction. Thus, we can rewrite (\ref{I-I}) with
the use of Eq. (\ref{J0}) as 
\begin{equation}
P_{n0}=-\gamma P_{F}+P_{n0}(\gamma +P_{n0}P_{F})\left(
I_{S}w/L_{s}-I_{pn}\right) I_{FS}^{-1}.  \label{P0}
\end{equation}
The current in FM-S junction $I_{FS}$ approaches a maximal value $%
I_{m}=J_{m}S$ at rather small bias, since $\gamma \simeq 1$ at $qV_{FS}>2T$ (%
\ref{TC}). When $I_{pn}\ll I_{FS}\simeq I_{m}$ and $I_{m}\gg I_{S}w/L_{s},$%
we get from Eq. (\ref{P0}) $P_{n0}\simeq -P_{F}.$

The way to maximize $P_{F}$ is evident from Eq.(\ref{PF}) showing that it
depends on $v_{\sigma 0}=v_{\sigma }(\Delta _{0}+qV_{FS})$ and $v_{t0}=\sqrt{%
2(\Delta -\Delta _{0}-qV_{FS})/m_{\ast }}$, i.e. it corresponds to an
electron energy $E=E_{c0}+qV_{FS}>F$, Fig.1. Therefore, $P_{F}$ depends on
bias voltage $V_{FS}$ of FM-S junction and one may be able to maximize a
spin extraction and accumulation by adjusting $V_{FS}$. The maximal $\left|
P_{n0}\right| $ can be achieved for the process of electron tunneling
through the $\delta $-doped layer when the bottom of conduction band in a
semiconductor $E_{c}=F+\Delta _{0}+qV_{FS}$ is close to a peak in the
density of states of minority carriers in the elemental ferromagnet, Fig.
1(c, curve g). Indeed, the density of states $g_{\downarrow }\propto
v_{\downarrow 0}^{-1}$ for minority d-electrons in Fe, Co, and Ni at $%
E=E_{F}+\Delta _{\downarrow }$ ($\Delta _{\downarrow }\simeq 0.1$ eV) is
much larger than that of the majority $d-$electrons, $g_{\uparrow }$, and $%
s- $electrons, $g_{s}$ \cite{Mor}. Therefore, due to $g_{\downarrow }\gg
g_{\uparrow }\gg $ $g_{s}$\ the polarization\ $\left| P_{n0}\right| \simeq
P_{F}$ may be close to 100\%.

Polarization of injection luminescence is determined by recombination of
injected spin polarized electrons and holes inside the $n^{\prime }$-region
of the double $p-n^{\prime }-n$ heterostructure of thickness $d\ll L_{s}$,
Fig. 1. The injection current of electrons with $\sigma =\uparrow $ into the 
$n^{\prime }$-region is 
\begin{equation}
I_{\uparrow pn}=I_{pn}(1+P_{n0})/2=q\delta n_{\uparrow }^{\prime }d/\tau
_{s}^{\prime }+qn_{\uparrow }^{\prime }d/\tau _{n},  \label{Iw}
\end{equation}
where $n_{\uparrow ,\downarrow }^{\prime }=n^{\prime }/2\pm \delta
n_{\uparrow }^{\prime }$ is the density of electrons with spin $\sigma
=\uparrow ,\downarrow $, $n^{\prime }=n_{\uparrow }^{\prime }+n_{\downarrow
}^{\prime }$, $\tau _{s}^{\prime }$ ($\tau _{n})$ is the time of spin
relaxation (recombination) of electrons in the $n^{\prime }$-region. We
notice that in III-V compound semiconductors, including GaAs and GaAlAs,
there are light holes with spin $\sigma =\pm 1/2$ and heavy holes with spin $%
\sigma =\pm 3/2$. Recombination of light and heavy holes with electrons
having spin $\sigma =1/2$ results in radiation of photons with an opposite
polarization ($l=-1)$. To exclude this undesirable effect, the thickness $d$
and the hole potential well in the narrow-gap $n^{\prime }$-region, Fig. 1,
should be such that the light holes, unlike the heavy holes, cannot be
localized inside the well. In this case we can neglect the light holes,
assuming that the concentration of heavy holes in the well is much larger
than that of the light holes. Then, the rate of polarized radiation
recombination is $R_{\sigma }=qn_{\sigma }^{\prime }d/\tau _{R}$ and the
polarization of radiation is $p=(R_{\uparrow }-R_{\downarrow })/(R_{\uparrow
}+R_{\downarrow })=2\delta n_{\uparrow }^{\prime }/n^{\prime }$. Since $%
I_{pn}=qn^{\prime }d/\tau _{n},$ we find from (\ref{Iw}) that $2\delta
n_{\uparrow }^{\prime }/n^{\prime }=P_{n0}\tau _{s}^{\prime }(\tau
_{s}^{\prime }+\tau _{n})^{-1}$, so that $p=P_{n0}\tau _{s}^{\prime }(\tau
_{s}^{\prime }+\tau _{n})^{-1}$. Thus, the radiation polarization $p$ can
approach maximum $p\simeq \left| P_{F}\right| $ at large current $I\simeq
I_{m}$ when $\tau <\tau _{s}^{\prime }$. The latter condition can be met at
high concentration $n^{\prime }$ when the time of radiation recombination $%
\tau _{R}$ $\simeq \tau _{n}<\tau _{s}^{\prime }$. For example, in GaAs $%
\tau _{R}\simeq 3\times 10^{-10}$s at $n\gtrsim 5\times 10^{17}$cm$^{-3}$ 
\cite{LO} and $\tau _{s}^{\prime }$ can be larger than $\tau _{R}$ \cite
{Wolf,Aw}.

We emphasize that spin polarization of {\em current} near a forward-biased
FM-S junction is very small. Indeed, according to Ref. \cite{BOB} $%
P_{j0}=(I_{\uparrow 0}-I_{\downarrow 0})/(I_{\uparrow 0}+I_{\downarrow
0})=P_{F}LI_{s}/L_{s}I_{m}\ll P_{F}$ at $I_{m}\gg I_{s}$ as $L<L_{s}$ Thus, 
{\em polarization of the recombination radiation would be high, }$p=\left|
P_{F}\right| $,{\em \ while the spin polarization of current in FM-S
junction is low.}
\begin{figure}[t]
\epsfxsize=3in 
\epsffile{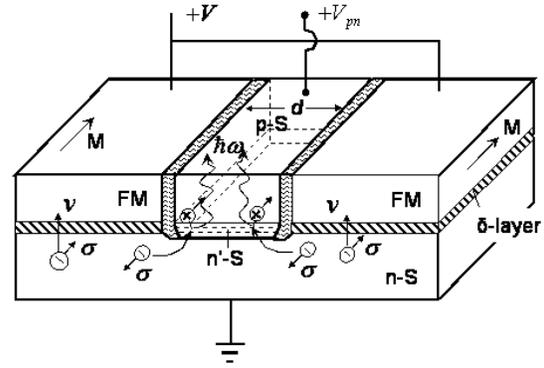}
\caption{Layout of the structure from Fig.~1, including FM layers and
semiconductor $n$- and $p$-regions. Here $n^{\prime }$ is made of a narrower
gap semiconductor, $\protect\delta -$doped layers are between FM layers and $%
n$-semiconductor. FM layers are separated by thin dielectic layers from $p$%
-region.}
\label{fig:fig2}
\end{figure}

Practical structures may have various layouts, with one example shown in
Fig.~2. It is clear that the distribution of $\delta n_{\uparrow }(r)$ in
such a two-dimensional structure is characterized by the length $L\lesssim
L_{s}$ in the direction $x$ where the electrical field $E$ can be strong,
and by{\bf \ }the diffusion length{\bf \ }$L_{s}$ in the ($y,z$) plane where
the field is weak. Therefore, the spin density near FM and $p-n$ junctions
will be close to $\delta n_{\uparrow 0}$ when the size of $p$-region is $%
d<L_{s}$. Thus, the above derivation and results for one-dimensional
structure, Fig.~1, are also valid for more complex geometry shown in Fig.~2.
The predicted effect should also exist for a reverse-biased FM-S junction
where the radiation polarization $p$ can approach $+P_{F}.$

\end{document}